\documentclass[12pt,fleqn]{article}
\usepackage[DIV12]{typearea}
\usepackage{amsmath,amsfonts,amssymb}
\usepackage[fleqn]{mathtools}
\usepackage{graphicx}
\usepackage{cite}
\usepackage{mathrsfs}
\usepackage{bbm}
\usepackage{pifont}
\usepackage{units}
\usepackage[small,loose]{subfigure}  % subfigures with (a), (b) etc., ... and own caption
\usepackage{xcolor}
\usepackage{pstricks,pst-node}

\usepackage[bookmarks=false]{hyperref}%\usepackage{hyperref}

\newcommand{\Eqref}[1]{equation~\eqref{#1}}
\newcommand{\Figref}[1]{figure~\ref{#1}}
\newcommand{\Tabref}[1]{table~\ref{#1}}

\newcommand{\eVdist}{\kern-0.06em}
\newcommand{\Ev}{\text{e\eVdist V}}     % solely as unit

\newcommand{\ev}{\:\text{e\eVdist V}}   % along with a number

\DeclareMathOperator{\diag}{diag}

\newcommand{\D}{\mathrm{d}}
\newcommand{\I}{\mathrm{i}}

\newcommand{\Z}[1]{\ensuremath{\mathbbm{Z}_{#1}}} % Z_N ->\Z{N}
\newcommand{\A}[1]{\ensuremath{\mathrm{A}_{#1}}}

\newcommand{\AfourFlavonA}{\ensuremath{\Phi_\nu}}
\newcommand{\AfourFlavonB}{\ensuremath{\Phi_e}}

\newcommand{\rep}[1]{\ensuremath{\boldsymbol{#1}}}

\allowdisplaybreaks[1]

\numberwithin{equation}{section}
\numberwithin{table}{section}

\def\mytitle{On predictions from spontaneously broken flavor symmetries}
\title{\mytitle}

\begin{document}

\begin{titlepage}

\begin{flushright}
 UCI-TR-2012-10\\
 TUM-HEP 849/12\\
 FLAVOUR(267104)-ERC-21\\
 CETUP*-12/008
\end{flushright}

\vspace*{1.0cm}

\begin{center}
{\Huge\bf
\mytitle
}

\vspace{1cm}

\textbf{Mu--Chun Chen\footnote[1]{Email: \texttt{muchunc@uci.edu}}{}}
\\[3mm]
\textit{\small
Department of Physics and Astronomy, University of California,\\
~~Irvine, California 92697--4575, USA
}
\\[5mm]
\textbf{
Maximilian Fallbacher\footnote[2]{Email:
\texttt{maximilian.fallbacher@ph.tum.de}}{},
Michael Ratz\footnote[3]{Email: \texttt{michael.ratz@tum.de}}{},
Christian Staudt\footnote[4]{Email: \texttt{christian.staudt@ph.tum.de}}{}
}
\\[3mm]
\textit{\small
Physik Department T30, Technische Universit\"at M\"unchen, \\
~~James--Franck--Stra\ss e, 85748 Garching, Germany
}
\end{center}

\vspace{1cm}

\begin{abstract}
We discuss the predictive power of supersymmetric models with flavor symmetries,
focusing on the lepton sector of the standard model. In particular, we comment
on schemes in which, after certain `flavons' acquire their vacuum expectation
values (VEVs), the charged lepton Yukawa couplings and the neutrino mass matrix
appear to have certain residual symmetries. In most analyses, only corrections
to the holomorphic superpotential from higher--dimensional operators are
considered (for instance, in order to generate a realistic $\theta_{13}$ mixing
angle). In general, however, the flavon VEVs also modify the K\"ahler potential
and, therefore, the model predictions. We show that these corrections to the naive
results can be sizable. Furthermore, we present simple analytic formulae
that allow us to understand the impact of these corrections on the predictions
for the masses and mixing parameters.
\end{abstract}

\end{titlepage}

\section{Introduction}

The observed patterns of fermion masses and mixing may originate from
underlying flavor symmetries. Typically, such flavor symmetries are assumed to
be spontaneously broken by the vacuum expectation values (VEVs) of certain
`flavon' fields. Given a large enough flavor symmetry, one may thus hope to
obtain a scheme that allows us to derive testable predictions. This applies, in
particular, to settings in which flavor is generated at a very high scale, which
cannot be directly accessed at colliders. 

In this work, we study supersymmetric extensions of the standard model, in which
flavor is generated at a high scale. For concreteness,  we will take the scale
of the flavon VEVs and the cut--off of the theory to be around the unification
scale, though our results do not depend on this choice. On the other hand, one
can imagine models in which there is a large difference between these two scales
or which are renormalizable. In such models, non--renormalizable corrections
including  the corrections from the K\"ahler potential discussed in this letter
become unimportant.

\begin{figure}
  \centering
  \includegraphics{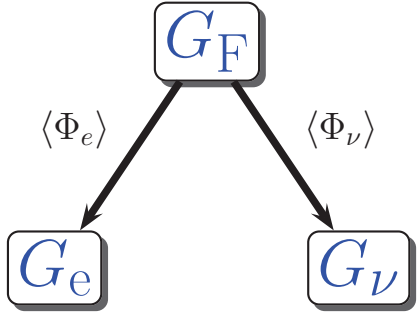}
  \caption{The flavor symmetry $G_\mathrm{F}$ gets broken to different subgroups in
   different sectors of the theory.}
\label{fig:VEVmisalignment}
\end{figure}

In order to be specific, we focus on the lepton sector of the theory, although
our analysis can also be applied to the quark sector. Generically, the relevant
superpotential reads, at the leading order,
\begin{equation}\label{eq:leadingW}
 \mathscr{W}_\mathrm{leading}
 ~=~
 \frac{1}{\Lambda}(\AfourFlavonB)_{gf}\,L^g\,R^f\,H_d
 +\frac{1}{\Lambda\,\Lambda_\nu}(\AfourFlavonA)_{gf}\,L^g\,H_u\,L^f\,H_u\;,
\end{equation}
where $L^g$ and $R^f$ (with the flavor indices $1\le f,g\le 3$) denote the
lepton doublets and singlets, respectively, $H_u$ and $H_d$ are the Higgs
doublets of the supersymmetric standard model, whereas $\Phi_e$ and $\Phi_\nu$
are the appropriate flavons. The two scales involved are the cut--off scale of the theory $\Lambda$ and
the see--saw scale $\Lambda_\nu$. Once $\Phi_e$ and $\Phi_\nu$ acquire
their VEVs, this leads to the effective superpotential
\begin{equation}\label{eq:Weff}
 \mathscr{W}_\mathrm{eff}~=~(Y_e)_{gf}\,L^g\,R^f\,H_d
 +\frac{1}{4}\kappa_{gf}\,L^g\,H_u\,L^f\,H_u\;.
\end{equation}
In many models, one is left with a situation in which the flavon VEVs
$\langle\Phi_e\rangle$ and $\langle\Phi_\nu\rangle$ respect certain residual
symmetries, which are then dubbed symmetries of the charged lepton Yukawa
couplings or the neutrino mass matrix, respectively (cf.\
figure~\ref{fig:VEVmisalignment}). Predictions of such models
are then based on these symmetries. 

However, one may question if these are really robust predictions of the
respective models. In particular, while certain terms in the superpotential
appear to possess the aforementioned symmetries, the Lagrangean density often
exhibits no residual symmetry. In other words, the combined VEVs
$\langle\Phi_e\rangle$ and $\langle\Phi_\nu\rangle$ break the flavor symmetry
completely. Moreover, the so--called predictions are subject to quantum
corrections. For instance, the bi--maximal \cite{Vissani:1997pa,Barger:1998ta}
or tri--bi--maximal \cite{Harrison:2002er} mixing patterns are known not
to be invariant under the renormalization group. On the other hand, the
statements below \eqref{eq:Weff} do not single out a particular scale.
Therefore, one may wonder how such corrections can be consistent with the
statement that the charged lepton Yukawa couplings or the neutrino mass matrix
exhibit certain symmetries.

At the first glance, one may think that the corrections are related to possible
higher--order terms that have to be added to the leading order superpotential
\eqref{eq:leadingW}. However, it is rather straightforward to construct
models in which such higher--order corrections are absent to all orders. We
will discuss such examples in a future publication \cite{Chen:2012pr}.

The true solution to this puzzle is that models of the above type do not predict
exact relations such as (tri--)bi--maximal mixing  due to the presence of the
K\"ahler corrections induced by the flavon
VEVs~\cite{Leurer:1993gy,Dudas:1995yu}, even if higher order holomorphic
corrections are absent.  
The K\"ahler potential should contain all terms
consistent with the flavor symmetry,
\begin{equation}
 K~=~K_\mathrm{canonical}+\Delta K\;,
\end{equation}
where the relevant canonical terms include (with the SM gauge multiplets 
being set to zero)
\begin{equation}
 K_\mathrm{canonical}~\supset~
 \left(L^f\right)^\dagger\,\delta_{fg}\,L^g
 +\left(R^f\right)^\dagger\,\delta_{fg}\,R^g
 \;,
\end{equation}
and $\Delta K$ contains contractions of $L^f$ and $R^f$ and their Hermitean
conjugates with the flavons. First of all, each of these terms in $\Delta K$
introduces one new parameter, i.e.\ its respective K\"ahler coefficient.
Furthermore, once the flavons attain their VEVs, the flavor symmetry is broken
thus modifying the K\"ahler metric. This modification $\Delta K$ of the K\"ahler
potential can be written as
\begin{equation}
  \Delta K~=~
  \left(L^f\right)^\dagger \, (\Delta K_L)_{fg} \,L^g 
  + \left(R^f\right)^\dagger \, (\Delta K_R)_{fg} \,R^g
  \;,
\end{equation}
with Hermitean matrices $\Delta K_{L}$ and $\Delta K_{R}$ whose structures are
determined by the flavor symmetries and the flavon VEVs.

The necessary field redefinitions to compensate for these additional terms and
to retrieve a canonical K\"ahler potential affect the superpotential. In
particular, the Majorana mass matrix of the neutrinos and the Yukawa coupling
matrix of the charged leptons are altered. This leads to changes of the neutrino
mixing parameters irrespective of the existence of higher--order terms in the
superpotential.

The purpose of this letter is to provide an analytic discussion of these
changes, using similar methods as for the renormalization group (RG) equations
in~\cite{Antusch:2003kp,Antusch:2005gp} and for the K\"ahler corrections
in~\cite{Antusch:2007ib,Antusch:2007vw}. We will argue that corrections from
these changes can be sizable and, therefore, without a better understanding of
the K\"ahler potential, predictions from spontaneously broken flavor symmetries
are incomplete. We leave a more detailed analysis for a future
publication~\cite{Chen:2012pr}. 

In section~\ref{sec:HolomorphicTerms} we will start out by describing a
well--known model that aims to explain the lepton mixing only with terms coming
from the superpotential. Section~\ref{sec:Kaehler} is then devoted to the
discussion of the K\"ahler corrections. Based on the results obtained in
sections~\ref{sec:linear} and \ref{sec:quadratic} and using analytic formulae
presented in section~\ref{sec:analytic}, we will argue in
section~\ref{sec:implications} that the changes compared to an analysis without
K\"ahler corrections are substantial. Finally, section~\ref{sec:Conclusions} summarizes our
conclusions.

\section{Predictions from the superpotential couplings}
\label{sec:HolomorphicTerms}

We first focus on the predictions of flavor models from the holomorphic
couplings of the theory, i.e.\ the superpotential. To be specific, we base 
our discussion on an example model \cite{Altarelli:2005yx} with an \A4 flavor symmetry~\cite{Ma:2004zv}, which serves as 
a prototype setting leading to tri--bi--maximal lepton mixing.

Since the following discussion heavily depends on the group structure of \A4, we
first review the necessary facts. In particular, these are the possible contractions
of fields transforming under this symmetry. \A4 has four inequivalent irreducible representations: 
three one--dimensional representations, denoted by \rep{1}, 
\rep{1'} and \rep{1''}, and one triplet, denoted by \rep{3}.
The relevant multiplication law is
\begin{equation}\label{eq:3x3}
 \rep{3}\otimes\rep{3}~=~\rep{1} \oplus \rep{1'} \oplus \rep{1''} \oplus \rep{3}_\mathrm{s} \oplus \rep{3}_\mathrm{a}\;,
\end{equation}
where $\rep{3}_\mathrm{s}$ and $\rep{3}_\mathrm{a}$ denote the symmetric and
the antisymmetric triplet combinations, respectively.
In terms of the components of the two triplets, $\rep{a}$ and $\rep{b}$,
\begin{subequations}
\label{eq:leptonReps}
\begin{eqnarray}
 \left(\rep{a} \otimes \rep{b} \right)_{\rep{1}}
 & = & 
 a_{1}\,b_{1}+a_{2}\,b_{3}+a_{3}\,b_{2}\;,
 \\ 
 \left(\rep{a}\otimes \rep{b} \right)_{\rep{1'}}
 & = & 
 a_{3}\,b_{3}+a_{1}\,b_{2}+a_{2}\,b_{1}\;, \\
 \left(\rep{a} \otimes \rep{b} \right)_{\rep{1''}}
 & = & 
 a_{2}\,b_{2}+a_{1}\,b_{3}+a_{3}\,b_{1}\;, \\
\left(\rep{a} \otimes \rep{b} \right)_{\rep{3}_\mathrm{s}}
 & = & \frac{1}{\sqrt{2}}\,
\left(\begin{array}{c}
 2 a_{1}\,b_{1} - a_{2}\,b_{3} - a_{3}\,b_{2}\\
 2 a_{3}\,b_{3} - a_{1}\,b_{2} - a_{2}\,b_{1}\\
 2 a_{2}\,b_{2} - a_{1}\,b_{3} - a_{3}\,b_{1}
\end{array}\right)\;, 
\\
 \left( \rep{a} \otimes \rep{b} \right)_{\rep{3}_\mathrm{a}}
 & = & \I\,\sqrt{\frac{3}{2}}\,
\left(\begin{array}{c}
 a_{2}\,b_{3} - a_{3}\,b_{2}\\
 a_{1}\,b_{2} - a_{2}\,b_{1}\\
 a_{3}\,b_{1} - a_{1}\,b_{3}
\end{array}\right)\;, 
\end{eqnarray}
\end{subequations}
where $\left(\rep{a} \otimes \rep{b} \right)_{\rep{R}}$ indicates that $\rep{a}$ and $\rep{b}$ are contracted to 
the representation \rep{R}. Note that there are different conventions for
normalizing the triplets $\rep{3}_i$ in the literature, and
the corresponding factors can be absorbed in the K\"ahler coefficients.

A well--known example for an \A4 tri--bi--maximal model is given by Altarelli et
al.\ \cite{Altarelli:2005yx}.
In this model, under \A4 the three generations of left--handed lepton doublets
transform as a triplet,  $L \sim$ \rep{3}, the right--handed charged leptons,
$e_\mathrm{R}$, $\mu_\mathrm{R}$ and $\tau_\mathrm{R}$, transform as 
\rep{1}, \rep{1''}, and \rep{1'}, respectively, 
and the Higgs fields $H_{u}$ and $H_{d}$ transform as pure singlets \rep{1}. Tri--bi--maximal
mixing is achieved by the introduction of three flavons: 
$\AfourFlavonA$ and $\AfourFlavonB$, 
both of which transform as triplets under the \A4 symmetry, and a pure \A4 singlet
$\xi \sim$ \rep{1}.  The couplings of the flavons to the SM fields are 
(cf.\ \Eqref{eq:leadingW})
\begin{eqnarray}
 \mathscr{W}_\nu
 & = &
 \frac{\lambda_1}{\Lambda\,\Lambda_{\mathrm{\nu}}}\,\left\{\left[(L\,H_u) \otimes
(L\,H_u)\right]_{\rep{3}_\mathrm{s}} \otimes
\AfourFlavonA\right\}_{\rep{1}}
 +
 \frac{\lambda_2}{\Lambda\,\Lambda_{\mathrm{\nu}}}\,\left[(L\,H_u) \otimes (L\,H_u)\right]_{\rep{1}}\,\xi\;,
 \\
 \mathscr{W}_e
 & = &
\frac{h_e}{\Lambda}\,\left(\AfourFlavonB\otimes L\right)_{\rep{1}}\,H_d\,e_\mathrm{R}
 +
 \frac{h_\mu}{\Lambda}\,\left(\AfourFlavonB\otimes L\right)_{\rep{1'}}\,H_d\,\mu_\mathrm{R}
 +
 \frac{h_\tau}{\Lambda}\,\left(\AfourFlavonB\otimes L\right)_{\rep{1''}}\,H_d\,\tau_\mathrm{R}
 \;,
\end{eqnarray}
where, as before, $\Lambda$ and $\Lambda_{\nu}$ are the flavon scale and
the see--saw scale, respectively.

Furthermore, the model is assumed to be invariant under a $\Z4$ symmetry under
which  $\AfourFlavonA$ and $\xi$ change sign, whereas $\AfourFlavonB$ is not
charged under this symmetry. The transformation properties of the leptons under this \Z4 are given by $L \rightarrow \I L$
and \mbox{$R \rightarrow - \I R$}.

The \A4 symmetry is then broken by assigning the following VEVs to the
flavons. 
\begin{subequations}\label{eq:A4VEVpattern}
\begin{eqnarray}
\langle\AfourFlavonA\rangle & = & \left(v,v,v\right)\;, \\
\langle\AfourFlavonB\rangle & = & \left(v',0,0\right)\;, \\
\langle\xi\rangle & = & w\;.
\end{eqnarray}
\end{subequations}
The resulting charged lepton Yukawa matrix after electroweak symmetry breaking
is diagonal and reads
\begin{equation}
 m_{e} ~=~ v_{d}\, \diag\left(y_{e},\, y_{\mu},\, y_{\tau}\right)\;,
\end{equation}
where $v_{d}$ is the VEV of $H_d$ and $y_{e,\,\mu,\,\tau}=h_{e,\,\mu,\,\tau}\,\frac{v'}{\Lambda}$.
The neutrino mass matrix, however, is non--diagonal. It is given by
\begin{equation}\label{eq:mnu}
 m_{\nu} ~=~ 
 \begin{pmatrix}
                a + 2d & -d & -d \\
		-d & 2d & a -d \\
                -d & a - d & 2d
\end{pmatrix}\;,
\end{equation}
with $a=2 \lambda_{2}\,
\frac{v_{u}^{2}}{\Lambda_{\mathrm{\nu}}}\,\frac{w}{\Lambda}$ and 
$d= \sqrt{2}  \lambda_{1}\, \frac{v_{u}^{2}}{\Lambda_{\nu}}\,\frac{v}{\Lambda}$, 
where $v_{u}$
is the VEV of $H_u$. The neutrino mass matrix is diagonalized by the
tri--bi--maximal mixing matrix
\begin{equation}
 U_\mathrm{TBM} ~=~ \begin{pmatrix}
                \sqrt{\frac{2}{3}} & \frac{1}{\sqrt{3}} & 0 \\
		- \frac{1}{\sqrt{6}} & \frac{1}{\sqrt{3}} & - \frac{1}{\sqrt{2}} \\
	      - \frac{1}{\sqrt{6}} & \frac{1}{\sqrt{3}} &  \frac{1}{\sqrt{2}}
	      \end{pmatrix}\;.
\end{equation} 
Since the charged lepton Yukawa matrix is already diagonal, the mixing matrix $U_\mathrm{PMNS}$ is identical to $U_\mathrm{TBM}$ and the corresponding mixing angles are shown in \Tabref{tab:angles}.
\begin{table}[t]
\centering
\renewcommand{\arraystretch}{1.5}
\begin{tabular}{lccc}
 & $\theta_{12}$ & $\theta_{13}$ & $\theta_{23}$\\
 TBM prediction: & $\arctan{\left( \sqrt{0.5} \right)} \approx 35.3^\circ$ &  $0$ & $45^\circ$\\
 Best fit values $(\pm 1 \sigma)$: & 
 $\left(33.6^{+1.1}_{-1.0}\right)^\circ$ & 
 $\left(8.93^{+0.46}_{-0.48}\right)^\circ$ & 
 $\left(38.4^{+1.4}_{-1.2}\right)^\circ$\\
\end{tabular}
\caption{Tri--bi--maximal prediction for the neutrino mixing angles and best fit
values from the global fit by \cite{Fogli:2012ua}.}
\label{tab:angles}
\end{table}
In the same table, the best fit values from the global fit on neutrino mixing
parameters by \cite{Fogli:2012ua} are quoted. As one can see, the 
recent measurement of $\theta_{13}$ has revealed a huge deviation from the
predicted TBM value. In addition, the TBM prediction for $\theta_{23}$ also does not agree
well with the current global fit value, which indicates a sizable deviation from maximal mixing.

These deviations seem to be difficult to explain with corrections coming from
the superpotential only. On the other hand, they might originate
from the  flavon VEV--induced K\"ahler corrections as we shall see in the
following section.

\section{Corrections due to K\"ahler potential terms}
\label{sec:Kaehler}

As discussed in the introduction, apart from the canonical terms,  there may
exist extra terms in the K\"ahler potential  induced by the flavon VEVs.  In the
\A4 example model discussed above, these terms are contractions of the
left--handed lepton doublets, which transform as an \A4 triplet, with one or
several flavons. After the flavons acquire a VEV, these terms lead to a K\"ahler
metric with off--diagonal terms.  We shall sketch the computation for the \A4
example model, leaving the general derivation to \cite{Chen:2012pr}.

\subsection{Linear flavon corrections}
\label{sec:linear}

The leading order contributions are linear in the flavons. These linear terms
are only suppressed by one power of the ratio of the flavon VEV to the
fundamental scale of the theory. The contributions in the \A4 model discussed
above read schematically
\begin{equation}
 \Delta K_\mathrm{linear} ~=~
 \sum\limits_{i\, \in \{\mathrm{a},\mathrm{s}\}}
 \left(\frac{\kappa_{\AfourFlavonA}^{(i)}}{\Lambda}\,\Delta K_{L^\dagger\,(L \otimes \AfourFlavonA)_{\rep{3}_i}}^{(i)}
 + \frac{\kappa_{\AfourFlavonB}^{(i)}}{\Lambda}\,\Delta K_{L^\dagger\,(L \otimes \AfourFlavonB)_{\rep{3}_i}}^{(i)}\right)
 + \frac{\kappa_\xi}{\Lambda} \,\Delta K_{\xi L^\dagger L}
 +\text{h.c.}\;.
\end{equation}
However, it is easy to forbid  any of these terms, by introducing an additional
symmetry (such as the \Z4 symmetry in the example model) under which all flavons are charged. Hence, we do not consider the
linear flavon corrections any further but turn to contributions which are
quadratic in the flavons, and cannot be forbidden by any (conventional)
symmetry.

\subsection{Second order corrections}
\label{sec:quadratic}

The corrections to the K\"ahler metric which are second order in the flavon VEVs
can be divided into two classes. The first class consists of terms that are of
the form $(L\AfourFlavonA)^\dagger (L\AfourFlavonA)$ or
$(L\AfourFlavonB)^\dagger (L\AfourFlavonB)$, i.e.\ they are quadratic in one
specific flavon. As mentioned above, these cannot be forbidden by a (conventional) symmetry.
This is not true for the second class which consists of terms of the form
$(L\AfourFlavonA)^\dagger (L\AfourFlavonB)$, i.e.\ they are contractions involving two
different flavons. For the same reasons as in the linear case, the second class
is not considered here.

All corrections discussed here can thus be obtained from suitable contractions
of the terms $(L \otimes \AfourFlavonA)_{\rep{R}}^\dagger (L \otimes
\AfourFlavonA)_{\rep{R'}}$ and $(L \otimes \AfourFlavonB)_{\rep{R}}^\dagger (L
\otimes \AfourFlavonB)_{\rep{R'}}$ using the rules stated in
\eqref{eq:leptonReps}.  Although there are numerous possible contractions,
several of them give the same correction $\Delta K$ to the K\"ahler metric up to
the respective K\"ahler coefficient which is a complex number. All in all, there
are 5 different matrices which have to be considered. The first three matrices
\begin{subequations}
\label{eq:KaehlerBasis}
\begin{equation}
 P_\mathrm{I}~=~\diag(1,0,0)\;,\quad
 P_\mathrm{II}~=~\diag(0,1,0)
 \quad\text{and}\quad
 P_\mathrm{III}~=~\diag(0,0,1)
\label{eq:pmatrix1}
\end{equation}
come from contractions of $L$ with $\AfourFlavonB$. That is, their contribution
is proportional to $(v')^2$, where $v'$ is the size of the VEV of
$\AfourFlavonB$, $\langle\AfourFlavonB\rangle=(v',0,0)$. The remaining two
matrices,
\begin{equation}
 P_\mathrm{IV}~=~\begin{pmatrix}
 1 & 1 & 1 \\ 1 & 1 & 1\\ 1 & 1 & 1
 \end{pmatrix}
 \quad\text{and}\quad
 P_\mathrm{V}~=~\begin{pmatrix}
 0 & \I & -\I \\ -\I & 0 & \I\\ \I & -\I & 0
 \end{pmatrix}
 \;,
\label{eq:pmatrix2}
\end{equation}
\end{subequations}
are contributions due to $\AfourFlavonA$. Therefore, their contribution in the
K\"ahler potential is proportional to $v^2$ which is defined by
$\langle\AfourFlavonA\rangle=(v,v,v)$.

The third flavon $\xi$ does not yield any relevant contribution since it can
only give an overall normalization factor, which does not change the mixing
angles. Another way of understanding this is by observing that $\xi$ is not
a flavon in the strict sense as it transforms trivially under \A4, such that its
VEV does not break \A4.

Each of the corrections is suppressed by the cut--off scale $\Lambda$ to the
second power. Furthermore, each of the terms comes with its own K\"ahler
coefficient $\kappa_i$, which, in general, is complex. Adding the Hermitean
conjugate always cancels either the term with the real or the imaginary part of
$\kappa_i$. We arranged our matrices $P_i$ in a way that all the
coefficients can be chosen real. 
However, the values of the K\"ahler coefficients $\kappa_{i}$ are not
fixed by the symmetries of the model and, therefore, their presence introduces
additional continuous parameters. One may hope to be able to compute them in a
possible UV completion of the model. Generically, these higher order terms in
the K\"ahler potential can come from integrating out heavy modes that are
required to complete the model in the UV. Since one expects to have several of
such modes, whose couplings to the zero modes of the theory can moreover
be unsuppressed, and due to group theoretical factors, the K\"ahler
coefficients can be of the order unity or even larger.

Let us comment that the K\"ahler corrections will, in general, also be
important for the question of VEV alignment. That is, the scalar potential that
fixes the VEVs of the flavons at some desired pattern will also be subject
to these corrections, and one might expect deviations from the fully symmetric
structures (such as those specified in \eqref{eq:A4VEVpattern}). We plan to
discuss these issues in more detail in our follow--up paper \cite{Chen:2012pr}.

\subsection{Corrections from the right--handed leptons}

In principle, there are also contributions from the right--handed sector.
However, in the model discussed here, all right--handed charged leptons are \A4
singlets, and therefore, the corresponding K\"ahler corrections can be made
diagonal. More precisely, possible off--diagonal terms can easily be forbidden
by additional symmetries (cf.\ the discussion in \ref{sec:linear}). Since
our basis is chosen such that the original charged lepton Yukawa matrix is
diagonal, a diagonal redefinition of the right--handed leptons $R^{f}$ cannot
induce any off--diagonal terms in the Yukawa matrix. Hence, the transformed
Yukawa matrix is still diagonal, only the eigenvalues may be changed. This
implies that such a field redefinition does not have any influence on the
neutrino mixing matrix. In conclusion, the model can be modified such that the
corrections from the right--handed sector cannot change the mixing parameters,
and therefore, they are not discussed any further.

\subsection{Analytic formulae for K\"ahler corrections}
\label{sec:analytic}

It is possible to derive some simple analytic formulae for the change of the
mixing parameters due to small non--diagonal terms in the K\"ahler
potential.\footnote{We only discuss the neutrino sector here. The left--handed
and right--handed charged lepton sectors can be dealt with separately in a
similar manner. This will be discussed in a future publication~\cite{Chen:2012pr}.} Suppose that,
after the flavon fields attain their VEVs, the K\"ahler potential reads
\begin{equation}
  K~=~K_\mathrm{canonical}+\Delta K~=~L^\dagger \, (1-2x\, P) \, L
\end{equation}
with a Hermitean matrix $P$ and an infinitesimal expansion parameter
$x$. The K\"ahler metric is
diagonalized to first order in $x$ by the field redefinition
\begin{equation}
  L ~\rightarrow~ L'~=~(1-x\,P) \, L\;.
\end{equation}
This field redefinition affects the effective neutrino mass operator
$\kappa$ for the canonically
normalized left--handed doublets $L^{\prime\,f}$,
\begin{equation}
  \mathscr{W}_\nu~\simeq~
  \frac{1}{4}\,(L^{\prime\,f}H_u)^T\,
  \left[\kappa + x \, P^T\, \kappa + x \, \kappa\, P\right]_{gf} 
  \, L^{\prime\,g}H_u\;,
\end{equation}
where $\kappa\cdot v_u^2=2m_\nu$ with $m_\nu$ specified in \Eqref{eq:mnu}.
That is, the neutrino mass operator has effectively become $x$--dependent, 
and the resulting neutrino mass matrix depends on $x$ as
\begin{equation}
 m_\nu(x)~\simeq~m_\nu + x \, P^T\, m_\nu + x \, m_\nu\, P\;.
\end{equation}
This leads to the differential equation
\begin{equation}\label{eq:DiffEq4mnu}
  \frac{\D m_\nu}{\D x}~=~ P^T\, m_\nu + m_\nu\, P
\end{equation}
for the neutrino mass matrix, which holds locally at $x=0$. This equation has
the same structure as the one governing the RG evolution
of the mass operator. In \cite{Antusch:2003kp}, analytic formulae describing the
evolution of the mixing parameters have been derived. Using an analogous
procedure, one can compute the derivatives of the mixing parameters at
$x=0$.

To this end, one derives a differential equation for $U_\nu$ from
\Eqref{eq:DiffEq4mnu} by substituting $U_\nu^* D_\nu U_\nu^\dagger$ for $m_\nu$,
where $D_\nu$ is the diagonalized neutrino mass matrix. This equation can then
be used to determine the entries of $U_\nu^\dagger \, \D U_\nu/\D x$, which can
also be written in terms of the mixing angles and phases of the standard
parametrization of $U_\mathrm{PMNS}$. A similar procedure was also already
used in \cite{Antusch:2007vw} to compute K\"ahler corrections.

With the K\"ahler coefficients and the ratios of flavon VEVs and high scale
$\Lambda$ as input parameters, the resulting formulae can be used to predict the
change of the mixing parameters due to a non--trivial K\"ahler metric for not
too large deviations from the canonical one. The detailed derivation of these
formulae and a more thorough discussion of their implications are deferred to a
later publication \cite{Chen:2012pr}. Here, we only discuss some examples for
the case of the \A4 model described above.

Let us briefly comment on the relation of K\"ahler corrections and RG evolution
(cf.\ also \cite{Antusch:2007ib}). First of all, unlike RG corrections, the
K\"ahler corrections are not loop--suppressed. Furthermore, while they are
similar in structure, generally the K\"ahler corrections can be along different
directions. In particular, they are not restricted to the diagonal. For example,
in the model considered, the main RG correction is essentially along the
direction specified by the matrix $P_{\mathrm{III}}$ in \Eqref{eq:pmatrix1}. The
K\"ahler corrections, however, can be along any of the five directions in
\Eqref{eq:KaehlerBasis}. Which one(s) of these five directions dominate(s)
depends upon the UV completion of the model.

\subsection{Implications for the $\boldsymbol{\A4}$ example model}
\label{sec:implications}

With the analytic formulae whose derivation was sketched briefly in the
foregoing section, we can compute the K\"ahler corrections which arise in the
example model \cite{Altarelli:2005yx} discussed in
section~\ref{sec:HolomorphicTerms}.

The most interesting correction is due to the matrix $P_\mathrm{V}$ in
\Eqref{eq:KaehlerBasis}. It originates from the term $(L \otimes
\AfourFlavonA)_{\rep{3}_\mathrm{a}}^\dagger\, (L \otimes
\AfourFlavonA)_{\rep{3}_\mathrm{s}}+\text{h.c.}$ in the K\"ahler potential. Performing the
\A4 contractions carefully, one finds that the additional K\"ahler
potential term is given by
\begin{equation}
\label{eq:Kcorr}
  \Delta K
  ~=~
  \kappa_\mathrm{V} \cdot \frac{v^2}{\Lambda^2} \cdot 3 \sqrt{3} 
  \cdot (L^f)^\dagger \, (P_\mathrm{V})_{fg} \, (L^g)\;,
\end{equation}
where $\kappa_\mathrm{V}$ denotes the relevant K\"ahler coefficient. 

The analytic formula for the change of $\theta_{13}$ compared to the case of a
canonical K\"ahler potential reads
\begin{eqnarray}
\label{eq:theta13An}
      \notag \Delta \theta_{13} & = &\kappa_\mathrm{V} \cdot \frac{v^2}{\Lambda^2} \cdot 3 \sqrt{3} \cdot \frac{1}{\sqrt{2}} \left( \frac{2 m_1}{m_1+m_3} + \frac{m_e^2}{m_\mu^2-m_e^2} + \frac{m_e^2}{m_\tau^2-m_e^2}\right)\\
      & \simeq & \kappa_\mathrm{V} \cdot \frac{v^2}{\Lambda^2} \cdot 3\sqrt{6} ~ \frac{m_1}{m_1+m_3}\;,
\end{eqnarray}
where the $m_i$ are the neutrino masses. In the second line, the very
small contribution of the charged leptons has been neglected.

\begin{figure}
  \centering
  \includegraphics{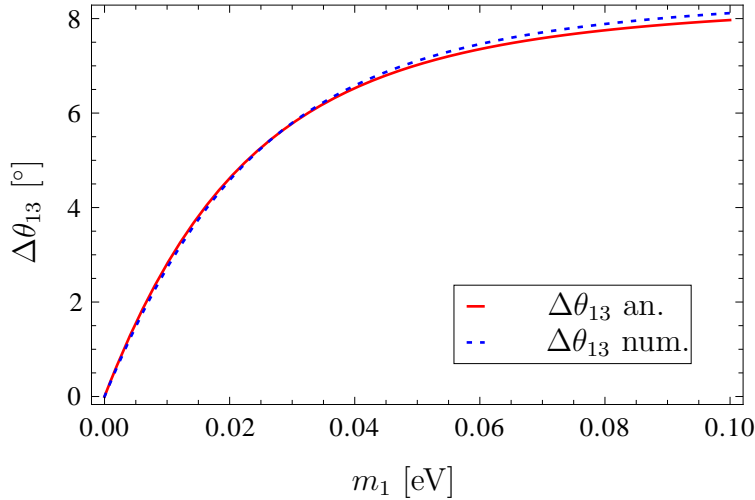}
  \caption{Change of $\theta_{13}$ due to the K\"ahler correction $\Delta K$
  shown in \Eqref{eq:Kcorr}   for $\kappa_\mathrm{V}\,v^2/\Lambda^2=(0.2)^2$.
  The continuous line shows the result of \Eqref{eq:theta13An}, which was
  obtained using a linear approximation (cf.\ section~\ref{sec:analytic}), while
  the dashed line shows the result of a numerical computation. As one can see,
  the linear approximation yields a very accurate estimate on the true change
  $\Delta\theta_{13}$.}
  \label{fig:theta13}
\end{figure}

In the following, we assume that the normal neutrino hierarchy is realised and
use the current PDG \cite{Beringer:1900zz} values for the differences of the
mass--squares,
\begin{equation}
  \Delta m_{21}^2 ~=~ 7.50 \cdot 10^{-5}\, (\Ev)^2
  \quad\text{and}\quad
  \Delta m_{32}^2 ~=~ 2.32 \cdot 10^{-3}\, (\Ev)^2\;,
\end{equation}
as input parameters. Moreover, the ratio of VEV to the fundamental scale
$v/\Lambda$ is set to $0.2$ and the K\"ahler coefficient $\kappa_\mathrm{V}$ is
set to 1. Then the variation of the change of $\theta_{13}$ with $m_1$ can be
studied and is shown in \Figref{fig:theta13}.
The deviation from the exact tri--bi--maximal prediction is substantial,
especially in the regime where $m_1$ gets large. This is also easy to see from
the analytic formula that asymptotically approaches a value of $\Delta
\theta_{13}\approx 8.42^\circ$ for $m_1 \rightarrow \infty$.  Based on the fact
that the differential equation for the K\"ahler corrections  is similar in
structure to the RG equation, our numerical result is consistent with the
expectation, as $m_1 \rightarrow  \mathcal{O}(0.1\ev)$ corresponds to the near
degenerate regime for the neutrino masses, where an enhanced correction to the
mixing angle is expected.

\begin{figure}
\centerline{
\subfigure[$\Delta\theta_{12}$.\label{fig:12}]{%
\includegraphics{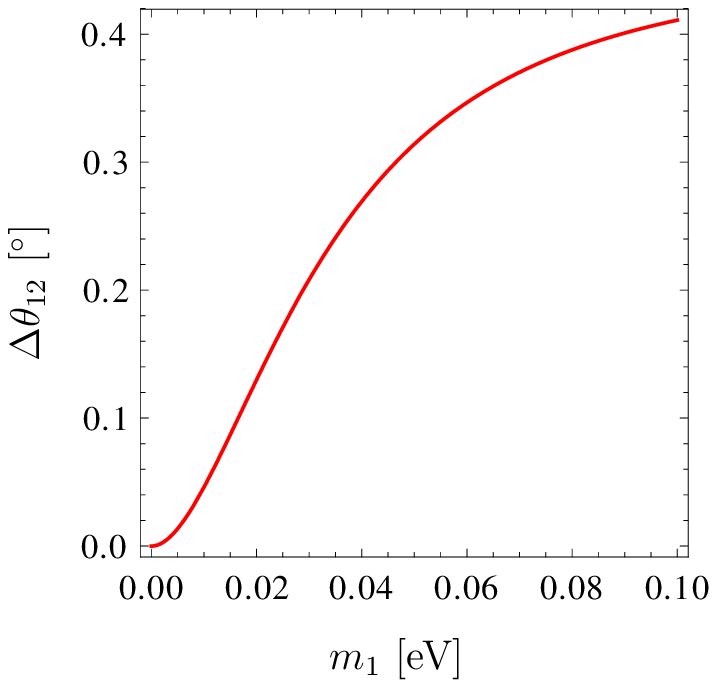}
}\hspace{0.9cm}
\subfigure[$\Delta\theta_{23}$.\label{fig:23}]{%
\includegraphics{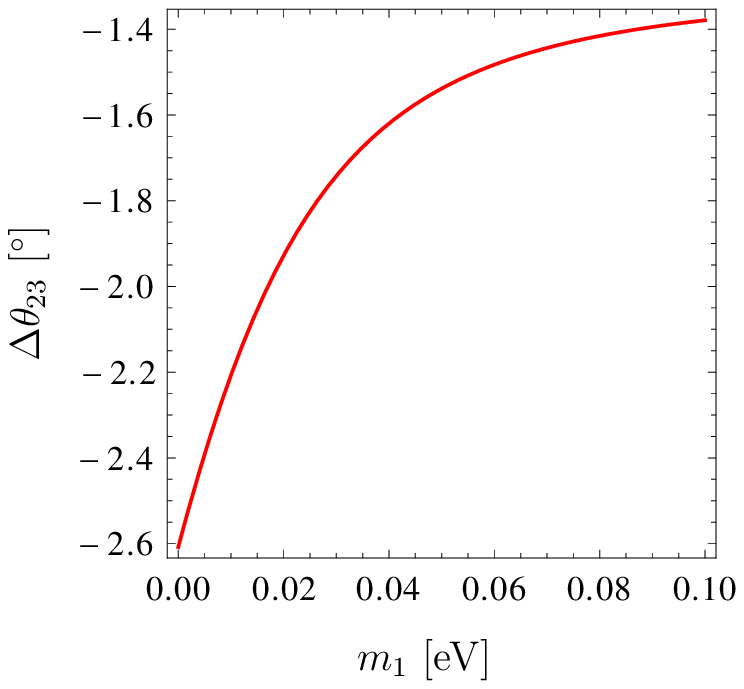}}
}
\caption{Changes of (a) $\theta_{12}$ and (b) $\theta_{23}$ due to 
  the K\"ahler correction $\Delta K$ shown in \Eqref{eq:Kcorr} for
  $\kappa_\mathrm{V}\, v^2/\Lambda^2 = (0.2)^2$ computed numerically.}
  \label{fig:theta1223}
\end{figure}

In contrast to the case of $\theta_{13}$, the changes of $\theta_{12}$ and
$\theta_{23}$ are predicted to be zero if one uses the linear extrapolation of
their changes starting from the tri--bi--maximal mixing pattern.  However, as we
have seen above, $\theta_{13}$ can undergo a substantial change such that also
the other two mixing angles change due to higher order non-linear terms. We have
confirmed this behavior numerically, using the MixingParameterTools
package \cite{Antusch:2005gp}. The dependence of the change on the lightest
neutrino mass $m_1$ is shown in figure~\ref{fig:theta1223}. Both changes are significantly smaller than the one of
$\theta_{13}$.

A further interesting consequence of the K\"ahler correction is the
generation of CP violation. It arises due to the fact that the matrix
$P_\mathrm{V}$ is complex. In fact, the Dirac CP phase $\delta$, which is not
properly defined for exact tri--bi--maximal mixing due to $\theta_{13}=0$, is
close to $\delta=3\pi /2$ taking into account the corrections. Note that
similar relations can also be obtained from the holomorphic superpotential
in models with $T'$ flavor symmetry \cite{Chen:2009gf}.

There can, of course, be additional contributions from other $P$ matrices.
Hence, as a second example of the implications of the K\"ahler corrections, we
discuss the case of $P_\mathrm{IV}$, which arises in all possible singlet
contractions of \AfourFlavonA, e.g.\ $(L \otimes
\AfourFlavonA)_{\rep{1}}^\dagger\, (L \otimes
\AfourFlavonA)_{\rep{1}}+\text{h.c}$. Including all coefficients, the
corresponding term in the K\"ahler potential is given by
\begin{equation}
\label{eq:Kcorr2}
  \Delta K
  ~=~
  \kappa_\mathrm{IV} \cdot \frac{v^2}{\Lambda^2} \cdot 2 
  \cdot (L^f)^\dagger \, (P_\mathrm{IV})_{fg} \, (L^g)\;.
\end{equation}
The resulting analytic formulae for the mixing angles are, in agreement with
the numerical computation, independent of the neutrino masses; they only
depend on the charged lepton masses, e.g.\
\begin{equation}
  \Delta \theta_{12}~=~ \kappa_\mathrm{IV} \cdot \frac{v^2}{\Lambda^2} \cdot \sqrt{2} \cdot \frac{\left(m_{\mu }^2\,m_{\tau }^2-m_e^4\right)}{\left(m_e^2-m_{\mu }^2\right) \left(m_e^2-m_{\tau }^2\right)}\;.
\end{equation}
$P_\mathrm{IV}$ does not induce any change of $\theta_{13}$, but the other two
mixing angles are shifted. In fact, the resulting change of $\theta_{12}$ for
$\kappa_\mathrm{IV}\, v^2/\Lambda^2=(0.2)^2$ is about $3.2^\circ$ while
the the change of $\theta_{23}$ is $-2.3^\circ$. In particular, for both
sign choices of $\kappa_\mathrm{IV}$ one of the two mixing angles is driven
further away from its best fit value.

The chosen examples illustrate that predictions which are solely based on
the inspection of the superpotential are not very reliable. Indeed, for example,
the global fit value for
$\theta_{13}=\left(8.93^{+0.46}_{-0.48}\right)^\circ$
\cite{Fogli:2012ua} (cf.\ table~\ref{tab:angles}) 
can be accommodated without resorting to higher--order
contributions from the superpotential, provided the neutrino mass spectrum
is not too hierarchical, the ratio of flavon VEV to the fundamental scale
$v/\Lambda$ is of the order of the Cabibbo angle and the K\"ahler coefficient
$\kappa_\mathrm{V}$ is of order one. On the other hand, there are K\"ahler
corrections that drive the theoretical predictions for the mixing parameters far
away from their current best fit values. Without any organizing principle for the
K\"ahler potential, it seems to be hardly possible to derive definite
predictions from discrete flavour symmetries.
Our results also show that the
K\"ahler corrections can be more significant than the effects of the RG
evolution.

\section{Conclusions}
\label{sec:Conclusions}

We have carefully re--examined models in which different flavons appear to break
a given flavor symmetry $G_\mathrm{F}$ down to different subgroups in different sectors of the
theory. In the context of supersymmetric settings, the fact that there is no
residual symmetry in the full Lagrangean manifests itself in corrections to the
K\"ahler potential $K$ that break $G_\mathrm{F}$ in all subsectors. We have argued
that the corresponding higher--order terms in $K$ are, in a way, unavoidable as
they cannot be forbidden by any (conventional) symmetry. These terms come with
certain coefficients, which are not determined by the symmetries of the model
and, therefore, introduce additional continuous parameters. We have also argued
that the K\"ahler corrections are generically much larger and, therefore, more
relevant than renormalization effects, which can also be understood as K\"ahler
corrections along a very specific direction. 

In order to make our analysis more concrete, we have outlined the discussion of
the corrections in a model based on the flavor symmetry $G_\mathrm{F}=\A4
\times\Z4$ \cite{Altarelli:2005yx}. We have presented results of an analytic
discussion of the K\"ahler corrections, i.e.\ simple analytic formulae
that allow us to express the change in the prediction on the mixing parameters
induced by the respective flavon VEVs. While leaving the full discussion for a 
future publication~\cite{Chen:2012pr}, we have explicitly shown that in the
simple $\A4$ model, which predicts tri--bi--maximal mixing at leading order, one
of the flavon VEVs induces a large variation of  the mixing angle $\theta_{13}$
while leaving the other mixing angles essentially unchanged. An optimistic
interpretation of this possibility may amount to the statement that even simple
models like the one discussed here can be consistent with the recent measurement
of $\theta_{13}$ \cite{Abe:2011fz,An:2012eh,Ahn:2012nd}. One the other hand, one
may be more critical and question the actual predictive power of a large class
of flavor models that exist in the literature. As we have seen in our second
example, K\"ahler corrections might significantly modify the predictions
of a model. Hence, one may actually argue that even in very simple models, a
better understanding of the K\"ahler potential is mandatory in order to achieve
an accuracy that can compete with the contemporary experimental precision.

In a future publication~\cite{Chen:2012pr}, we will provide more details on the derivation of the
analytic formulae used in this letter.

\subsection*{Acknowledgments}

M.-C.C. would like to thank TU M\"unchen, where part of the work was done, for
hospitality. M.R. would like to thank the UC Irvine, where part of this work was
done, for hospitality. This work was partially supported by the Deutsche
Forschungsgemeinschaft (DFG) through the cluster of excellence ``Origin and
Structure of the Universe'' and the Graduiertenkolleg ``Particle Physics at the
Energy Frontier of New Phenomena''. This research was done in the context of the
ERC Advanced Grant project ``FLAVOUR''~(267104), and was partially supported by
the U.S. National Science Foundation under Grant No.\ PHY-0970173. We thank the
Aspen Center for Physics, where this discussion was initiated,  the Galileo
Galilei Institute for Theoretical Physics (GGI),  the Simons Center for Geometry
and Physics in Stony Brook,  and the Center for Theoretical Underground Physics
and Related Areas (CETUP* 2012) in South Dakota for their hospitality and for
partial support during the completion of this work.

\bibliography{Orbifold}
\addcontentsline{toc}{section}{Bibliography}
\bibliographystyle{NewArXiv}
\end{document}